# Mechanics of Individual, Isolated Vortices in a Cuprate Superconductor

Ophir M. Auslaender,[1] Lan Luan,[1] Eric W. J. Straver,[1] Jennifer E. Hoffman,[1,2] Nicholas C. Koshnick,[1] Eli Zeldov,[1,3] Douglas A. Bonn,[4] Ruixing Liang,[4] Walter N. Hardy[4] & Kathryn A. Moler[1]

[1]*Geballe Laboratory for Advanced Materials, Stanford University, Stanford, CA 94305, USA;* [2]*Department of Physics, Harvard University, Cambridge, MA 02138, USA;* [3]*Department of Condensed Matter Physics, Weizmann Institute of Science, Rehovot 76100, Israel;* [4]*Department of Physics and Astronomy, University of British Columbia, Vancouver, BC, Canada V6T 1Z1.*

**Superconductors often contain quantized microscopic whirlpools of electrons, called vortices, that can be modeled as one-dimensional elastic objects.[1] Vortices are a diverse playground for condensed matter because of the interplay between thermal fluctuations, vortex-vortex interactions, and the interaction of the vortex core with the three-dimensional disorder landscape.[2-5] While vortex matter has been studied extensively,[1, 6, 7] the static and dynamic properties of an individual vortex have not. Here we employ magnetic force microscopy (MFM) to image and manipulate individual vortices in detwinned, single crystal $YBa_2Cu_3O_{6.991}$ (YBCO), directly measuring the interaction of a moving vortex with the local disorder potential. We find an unexpected and dramatic enhancement of the response of a vortex to pulling when we wiggle it transversely. In addition, we find enhanced vortex pinning anisotropy that suggests clustering of oxygen vacancies in our sample and demonstrates the power of MFM to probe vortex structure and microscopic defects that cause pinning.**

A superconducting vortex is characterized by two length-scales: the nanoscale core size and the much larger magnetic penetration depth, $\lambda$. Pinning can occur when the core is co-located with a defect that locally suppresses superconductivity. $\lambda$ determines the decay length for the currents encircling the core and the elastic properties of a vortex. Here we probe the dynamics of an individual driven vortex, which is especially interesting when pinning and elasticity compete. Furthermore, we establish a dragged vortex as a probe that extends deep into the bulk of the sample to interact with defects far from the surface, circumventing disadvantages of other local-probe techniques that give information only about the immediate vicinity of the surface.[8] Previous strategies for manipulating single vortices in superconductors usually applied forces relatively delocalized on the scale of $\lambda$.[9-11] We use MFM to combine imaging and vortex manipulation with a level of control far beyond what has been demonstrated before.[12, 13] This capability may enable testing vortex entanglement[14] and schemes for quantum computation.[15, 16]

Our sample is ideal for studying the interplay between pinning and elasticity, with weak, well-controlled, pinning and fairly rigid vortices. In YBCO superconductivity arises in $CuO_2$ planes, parallel to the a and b axes, and in Cu-O chains, along the b-axis. In pristine samples, such as ours, oxygen vacancies in the chains are the dominant source of pinning.[17-19] The orthorhombic crystal structure gives rise to penetration depth anisotropy, allowing us to determine the orientation of the crystal axes in situ [Supplementary Fig. (SFig.) 1]. We mounted the platelet shaped sample with the c-axis along $\hat{z}$ and the a-axis ~9° from $\hat{x}$ (cf. Fig. 1a).

MFM employs a sharp magnetic tip on a flexible cantilever. Our tip magnetization provided an attractive tip-vortex force, $\vec{F}$. While rastering in the $x-y$ plane, parallel to the sample surface, we measured local variations in the resonant frequency of the cantilever to determine $\partial F_z / \partial z$ .[20] We deliberately used the lateral components of $\vec{F}$ , $F_x \hat{x} + F_y \hat{y} \equiv \vec{F}_{lat}$ , for vortex manipulation.

At low temperature ($T \approx 5$K), we observed no vortex motion up to our largest lateral force, 20 pN. At higher temperatures, pinning is reduced, and we could reduce the tip-sample distance $z$ to tune from non-invasive imaging to manipulation. Figure 1 shows typical examples for individual, well isolated, vortices. The similarity of the behaviour of vortices pinned at different locations indicates the uniformity of the pinning landscape in this sample.

Figure 1d shows one of our two major findings: "vortex wiggling". Adding an alternating transverse force enhances vortex dragging dramatically. The wiggling occurs because of the rastering associated with imaging: between incremental steps forward along the "slow" scan direction we raster the tip left and right along the "fast" scan axis. If $F_{lat} \equiv |\vec{F}_{lat}|$ is small this creates an image of a stationary vortex (Figs. 1a,c,e). When $F_{lat}$ is large, the vortex moves as the tip passes over it (Figs. 1b,d,f). While this motion is substantial in the quasi-one-dimensional (1D) scans along the fast axis (Fig. 1f), it is much larger along the slow axis (Fig. 1d).

We study wiggling further in Fig. 2. Figure 2a is a scan that we acquired after imaging the same area over and over, resulting in an enhanced wiggling effect. Line scans along $x$ from that image (Fig. 2b) do not differ from Fig. 1f. To study the vortex after completing the scan in Fig. 2a, we moved the tip back and forth over the centre of the vortex along $y$ to obtain Figs. 2c,d. Details of the motion along the fast axis (Fig. 2b) are similar to motion along what was the slow axis (Figs. 2c,d) - a vortex moves in jerks, reminiscent of avalanches[5] – but the total distance moved differs substantially. Moreover, Figs. 2c-d show that while a vortex moves very freely between the initial and final positions in Fig. 2a, it does not readily move outside this range. In fact, we were never able to permanently drag a vortex very far from its original location, contrary to thin films.[12] This tethering suggests that each vortex was pinned along its full length across the 40μm-thick crystal, and that we observed the vortex stretching.

A vortex presumably stops moving where elasticity and pinning balance $F_{lat}$. Qualitatively, wiggling helps segments of the vortex to depin, facilitating the extra motion along the slow axis. Confirming this is Fig. 2 with the enhancement of effect by adding wiggling cycles (further test in SFig. 2). Wiggling is reminiscent of "vortex shaking", used to accelerate equilibration in vortex matter by oscillating a magnetic field perpendicular to the applied magnetic field generating the vortices,[6, 21, 22] and may be the mechanism for it. The models presented below describe motion without wiggling, as we observe along the fast axis. Our single vortex data should be amenable to more advanced and quantitative theoretical modelling to describe the wiggling.

The behaviour along the fast axis can be analyzed as individual 1D scans. As the tip approaches a vortex, $F_{lat}$ increases until, if it overcomes pinning, the vortex moves. The vortex then moves until pinning and the growing elastic force balance $F_{lat}$, presumably when $F_{lat}$ is maximal, allowing us to treat this as a static problem. We plot the dependence of $w$, the distance dragged along the fast axis, on $F_{lat}^{\max} \equiv \max(F_{lat})$ in Fig. 3. We have constructed a model based on weak collective pinning (WCP) for a single vortex[1] to explain this data [Supplementary Discussion 2 (SD 2)]. WCP assumes that pinning is only by the collective, cumulative, effect of many pinning sites, each too weak to pin a vortex on its own, and that vortices are elastic strings,[23] as described in Ginzburg-Landau (GL) theory.[1] A characteristic length, $L_c$, emerges from the competition between pinning and elasticity: a vortex takes advantage of pinning by bending on a length $L \gg L_c$, but is too rigid to bend for $L \ll L_c$. In WCP the vortex is broken into elastically coupled $L_c$-segments, each pinned by a characteristic force, $F_p$.

We assume that each vortex is initially along the c-axis, that only the top portion interacts directly with the tip and that $L_c << \lambda_{ab}$ ($\lambda_{ab}$, the in-plane penetration depth). Each $L_c$-segment is subject to elastic forces from neighbouring segments and a pinning force up to $F_p$, modelled as static friction. We find that the top portion of a vortex depins when $F_{lat}^{\max} > F_p \Lambda / L_c$ ($\Lambda$ of order $\lambda_{ab}$) and that:

$$w \sim \frac{F_{lat}^{\max}\left(F_{lat}^{\max} - F_1\right)}{2 F_p \varepsilon_\perp / L_c}, \qquad (1)$$

for $F_{lat}^{\max} >> F_p \Lambda / L_c$ (SD 3). Here $F_1 / F_p \approx 2 \Lambda / L_c$ and $\varepsilon_\perp$ is the tilt modulus. Figure 3b shows a reasonable fit for large $F_{lat}^{\max}$, where the exact shape of the top portion of the vortex and the fine details of the pinning landscape are not important.

Figure 3b also highlights our second main finding – in-plane anisotropy due to the crystal: we can drag a vortex farther along the fast axis if it is along the YBCO b-axis than if it is along the a-axis. This is also apparent in Fig. 4. Figures 4c-d show the dependence of $w$ on the fast scan angle (cf. SFig. 3 for $w$ here). We found a similar effect using different tips [see Supplementary Table (ST)] and for every vortex that we probed.

Equation 1 explains the anisotropy qualitatively. The penetration depth imparts in-plane anisotropy to $\varepsilon_\perp$,[1, 24] implying that it is easier to tilt a vortex towards the b-axis. Additionally, because of the in-plane anisotropy of the vortex core radius, even for point pinning, the effective pinning potential is shallower along the b-axis, implying a smaller depinning force. Equation 1 thus naturally explains the weak dependence of the anisotropy on $F_{lat}^{\max}$ and $T$, seen in Figs. 4c-d. When $F_{lat}^{\max} >> F_1$ the only dependence on $F_{lat}^{\max}$ is an overall scale. The same is true for the $T$-dependence, because both the superconducting parameters and the defect structure depend only weakly on $T$ in our range.[25]

For quantitative analysis, we have determined the angular dependence of $w \propto L_c / \varepsilon_\perp F_p$ (SD 4), which depends on the anisotropy factor $\zeta$. Fitting all five data sets in Figs. 4c-d we find $\zeta = 1.6$ (SFig. 5), in clear disagreement with our direct measurement (SFig. 1) and the known value,[25] $\zeta = 1.3$. We surmise that there is an extra source of anisotropy.

A likely source for extra pinning anisotropy is nanoscale clustering of oxygen vacancies along the Cu-O chains[26, 27] (Fig. 4e). We modified our model (SD 5), imposing $\zeta = 1.3$, and fit the data in Figs. 4c-d for the cluster size. We find clusters of order 10 vacancies along the Cu-O chains (SD 5), large but reasonable for the sample's length of time at room temperature, where oxygen vacancies migrate slowly and cluster.[28] It is also possible that non-GL physics affects the core structure,[29] changing details of the pinning-force anisotropy. These results demonstrate that single vortex manipulation is a local probe of both the core and the defect structure on a scale down to the core size. For example, in samples without intrinsic a-b anisotropy and with known defect structure, MFM could be used to probe the intrinsic structure of the vortex core itself.

Despite the fact that YBCO is one of the most studied superconductors, our data reveal major surprises about the behaviour of individual vortices. A model based on weak collective pinning for a single vortex quantitatively describes quasistatic aspects of vortex motion with the incorporation of anisotropy in the local microscopic pinning. This demonstrates that single vortex manipulation is a local probe of the structure of both the vortex and the pinning defects. Further work is required to describe the dynamic aspects of individual vortex motion that we revealed: its stochastic nature and the dramatic effect of transverse wiggling. In particular: how do the dynamics alter the effective pinning landscape? How do they affect the mechanical properties? Practically, wiggling is an important tool for future experiments that require pulling vortices long distances, e.g. in the study of vortex entanglement.[14] Our results show the utility of local force probes for accessing the pinning properties and mechanical behaviour of individual vortices, whose collective behaviour is of great importance for the properties of superconductors.

## Methods Summary

**Frequency modulated MFM.[20]** We oscillate the cantilever along $z$ and measure the resonant frequency, $f_0$, which shifts by $\Delta f$ in a force gradient. Assuming small oscillation amplitude, $\partial F_z / \partial z \approx -2c\,\Delta f / f_0$ ($c =$ cantilever spring constant, cf. ST).

**Sample details.** The platelet-shaped single crystal was grown from flux in a $BaZrO_3$ crucible for high purity and crystallinity.[17] The (001) surfaces were free of visible inclusions. Mechanical detwinning was followed by annealing to oxygen content 7-$\delta$ = 6.991, implying $T_c \approx 88K$.[19] The sample was stored at room temperature for a few years. The finite a-b anisotropy allowed us to identify crystal orientation by determining the directions along which vortex spacings were extremal (SFig. 1). The ratio between these spacings is $\zeta = 1.3$, corroborating previous results on similar samples.[25]

Supplementary Information accompanies the paper on www.nature.com/nature.

Research supported by DOE Contract No. DE-AC02-76SF00515, AFOSR grant FA50-05-1-0290, the Packard Foundation and the US-Israel Binational Science Foundation (BSF). The experimental set up was developed with extensive help from D. Rugar. We acknowledge useful discussions with J. R. Kirtley, H. Bluhm, G. P. Mikitik, E. H. Brandt and V. B. Geshkenbein.

Correspondence and requests for materials should be addressed to O.M.A. (ophir@stanford.edu).

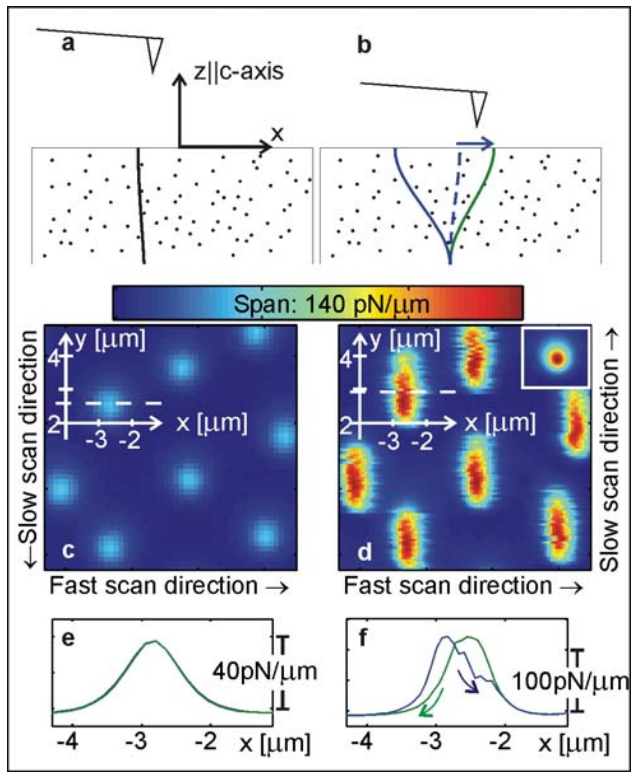

**Figure 1:** MFM imaging and manipulation of individual vortices in YBCO at $T = 22.3$K. **(a,b)** Cartoons, not drawn to scale, showing an MFM tip (triangles) that attracts a vortex (thick lines) in a sample with randomly distributed point pinning sites (dots). **(a)** At "surveillance" height the applied force $F_{lat}$ is too weak to move the vortex. **(b)** At manipulation height, the vortex moves right and then left, as the tip rasters over it. Here we illustrate what happens in a scan along $+\hat{x}$, as indicated by the arrow. Also shown are three configurations of the vortex, previously dragged along $-\hat{x}$: the blue line on the left illustrates the vortex before the tip drags it to the right, the dashed blue line shows an intermediate configuration as it follows the tip and the green line on the right shows the final configuration, after the tip moves away. **(c,d)** MFM scans for two different scan heights (color-scale gives $\partial F_z / \partial z$, fast scan and slow scan directions are indicated on the frame). **(c)** $z = 420$nm (maximum applied lateral force $F_{lat}^{max} \approx 5$pN), not low enough to perturb vortices at this temperature. **(d)** $z = 170$nm ($F_{lat}^{max} \approx 12$pN), low enough to drag the vortices significantly. **d - inset:** scan at 5.2K, showing a stationary vortex at tip height and force comparable to those in the main panel. **(e)** Line cut through the data in (c) along the dashed line, showing the signal from a stationary vortex. **(f)** Line cut through the data in (d) along the dashed line, showing a typical signal from a dragged vortex. Right arrow shows the data acquired with the tip moving along $+\hat{x}$, as in (d), the left arrow shows the data acquired with the tip moving back along $-\hat{x}$.

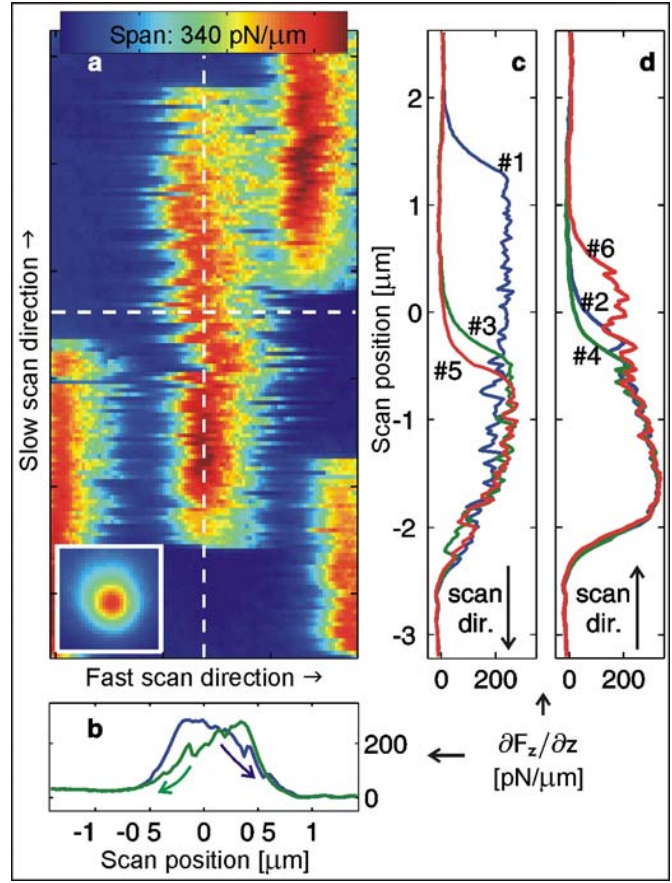

**Figure 2:** MFM image and line-scans at $T = 20$K, showing how wiggling enhances dragging along the slow direction. **(a)** Scan at $z = 80$nm ($F_{lat}^{max} \approx 20$pN), acquired after repeated imaging, reversing the slow scan direction for each new scan, which enhances the wiggling effect. Scan directions are denoted on the frame, dashed lines show the trajectories of the line-scans in (b-d). **Inset:** Scan with similar parameters at 5.2K, where vortices were immobile. **(b)** Line scans from (a) along the horizontal dashed line. Arrows show the scan direction. **(c,d)** Immediately after the scan in (a) we scanned back and forth along the vertical dashed line at $z = 80$ nm [(c) scans down, (d) scans up]. Indices: order of line-scan acquisition.

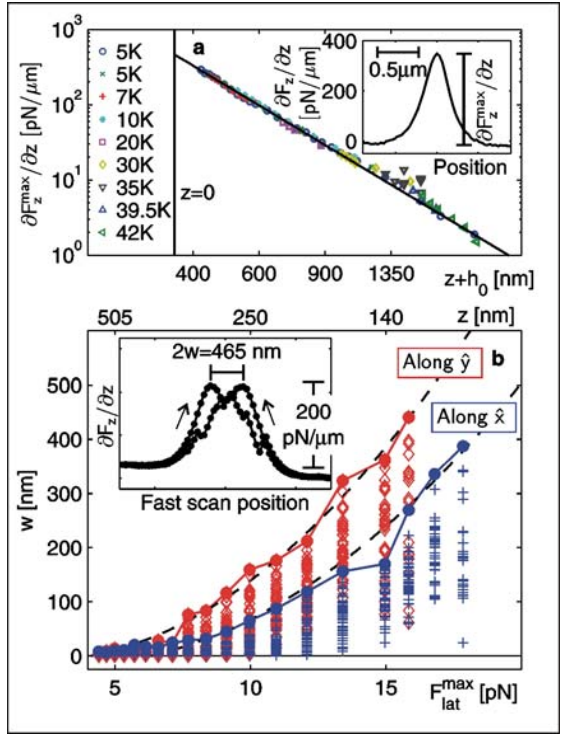

**Figure 3:** Force dependence of the dragged distance, $w$, along the fast axis. **(a)** Force calibration: magnitude of peaks from vortices, $\max(\partial F_z/\partial z) \equiv \partial F_z^{\max}/\partial z$, vs. scan height at various temperatures. Clearly the temperature dependence in this range is weak. The dashed line is a fit to $\max(\partial F_z/\partial z) = (\widetilde{m}\Phi_0/\pi)/(z+h_0)^3$ ($h_0 = 360\pm10$nm, $\widetilde{m} = 32\pm2$nAm, SD 1). The excellent fit allows us to use $F_z^{\max} = (\widetilde{m}\Phi_0/2\pi)/(z+h_0)^2$. The maximum applied lateral force is given by $F_{lat}^{\max} = \alpha F_z^{\max}$. For a wide range of tip shapes $0.3 < \alpha < 0.4$.[30] We set $\alpha = 0.35$, adding at most 25% systematic error to $F_{lat}^{\max}$.

**Inset:** Single line from a scan at $z = 65$nm, $T = 5.2$K, showing the peak height for an immobile vortex [here $\max(\partial F_z/\partial z) = 365$ pN/µm]. **(b)** Distance moved by the vortex along the fast direction vs. $z$ (top axis) acquired at $T = 25$K vs. $F_{lat}^{\max}$ (bottom axis). In addition to the maximum $w$ (filled circles), we plot the distribution (diamonds and crosses), which shows the stochasticity of the vortex motion. Other errors are not shown. Dashed lines are fits to Eq. 1 (along $\hat{x}: F_1 = 6.3\pm0.7$pN, $F_p \varepsilon_\perp/L_c = 270\pm20$pN²/µm; along $\hat{y}: F_1 = 4.1\pm1.4$pN, $F_p \varepsilon_\perp/L_c = 210\pm30$ pN²/µm). **Inset:** subsequent line scans showing the jerky nature of the vortex motion and how $w$ was extracted from the difference of the vortex positions in pairs of subsequent line scans.

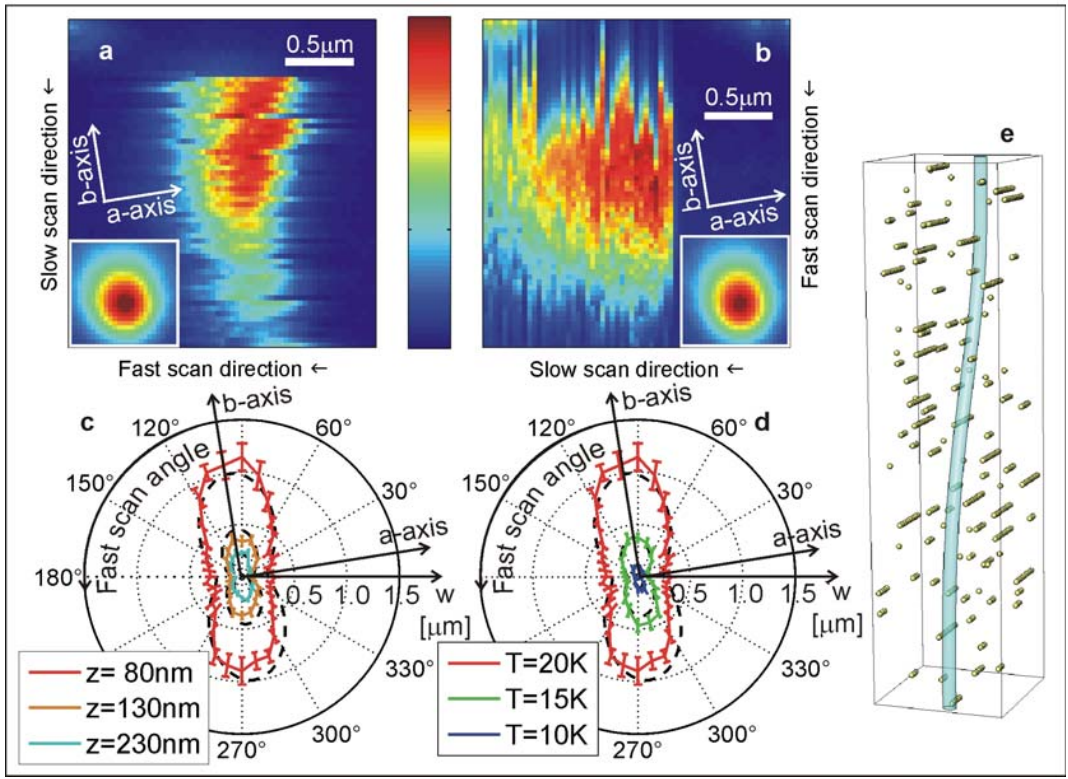

**Figure 4:** Images and analysis showing the anisotropy of dragging when the fast axis is along the a-axis and when it is along the b-axis. **(a,b)** Vortex motion when the fast scan is along $\hat{x}$ (a) and $\hat{y}$ (b) ($T = 20$K, $z = 80$nm, $F_{lat}^{max} \approx 20$pN). Note the erratic nature of the motion and the abrupt snap-in to the tip on approaching the vortex along the slow axis, apparent as a sharp onset of the signal. **Insets:** Images of immobile vortices obtained with the same tip, scan height and scan directions at $T = 5.2$K. **(c,d)** Distance moved along the fast axis, $w$, measured at the maximum lateral force, vs. scan angle measured from $\hat{x}$. Bars denote 70% confidence intervals. Dashed lines show the fit described in the text, with the oxygen vacancy cluster size $2R_b$ as a free parameter. Fit result: $2R_b / \xi_{ab} = 0.7$. **(c)** $T = 20$K at $z = 80$nm, 130nm, 230nm ($F_{lat}^{max} \approx 20$, 15, 10pN, respectively). **(d)** $z = 80$nm, for $T = 20$K, 15K, 10K. **(e)** Cartoon of vortex core meandering across the crystal in the presence of point defects clustered along the b-axis.

# Mechanics of Individual, Isolated Vortices in a Cuprate Superconductor

Ophir M. Auslaender, Lan Luan, Eric W. J. Straver, Jennifer E. Hoffman, Nicholas C. Koshnick, Eli Zeldov, Douglas A. Bonn, Ruixing Liang, Walter N. Hardy & Kathryn A. Moler

## Supplementary Information

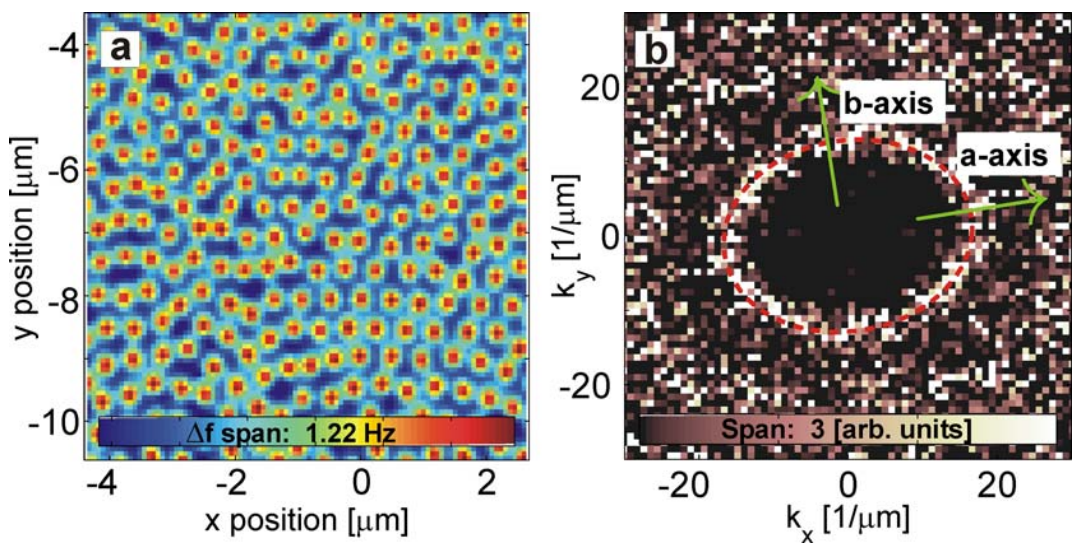

**Supplementary Figure 1:** Identification of the crystal axes and the determination of $\zeta = \lambda_a/\lambda_b$. **(A)** Vortices in the YBCO single crystal at $T = 6$K. Scan height $z = 85$nm. Flux density extracted from this image is 0.011T. In this scan there is a slight field gradient along $\hat{y}$, which does not affect the conclusions here (similar results were obtained for scans without any gradient). The tip-vortex interaction was repulsive in this scan, the only such scan here, or in the main text. **(B)** Fourier transform of the vortex locations. Clearly seen is an elliptic band, also highlighted by a dashed line, from which we determine that the a-axis and the b-axis are oriented 9° ± 4° to the scan directions, as indicated by the arrows. The eccentricity of the ellipse gives the in-plane anisotropy: $\zeta = \lambda_a/\lambda_b = 1.3$. This compares well with previous results.[1-6]

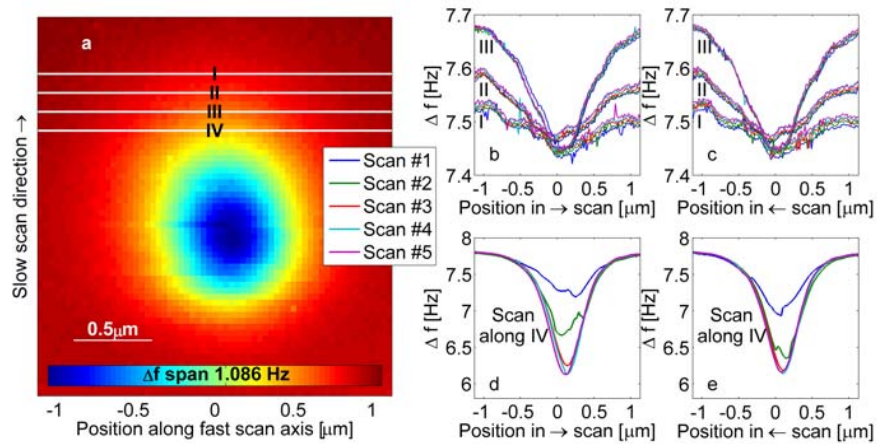

**Supplementary Figure 2:** Test of transverse wiggling ($T = 20$K). **(A)** Scan at $z = 210$nm ($F_{lat}^{max} \approx 12$pN). The vortex is seen to move under the influence of the tip, but not extensively. The lines marked by I-IV denote where and the order we scanned the tip back and forth after this scan. The legend enumerates the order of these lines scans (#1-#5), which are shown in panels (B-E). (Note slight tilt along the slow scan direction, which is a result of a slight angle between the scan plane and the surface of the sample.) **(B-E)** Line scans along the horizontal lines in (A) at $z = 105$nm ($F_{lat}^{max} \approx 17$pN). **(B)** Scans from left to right along tracks I-III. Each line scan was repeated five times, as denoted in the legend. The slight angle between the scan plane and the sample plane is evident in the different offset of the background value of $\Delta f$ for I, II & III. No motion is evident in scans I, II. Slight motion along the fast scan axis (~50nm) is evident in the offset between scan #1 and the rest of the scans in III. **(C)** Like B but for scans from right to left. No motion is evident. **(D, E)** Vortex motion towards the tip along both the slow direction and the fast direction are evident. Clearly every time the tip approaches the point of closest approach, the vortex moves a little towards it. One can see that in the first scan the vortex was near where it was after the last scan in (C) and that it progressed towards the tip in subsequent line scans.

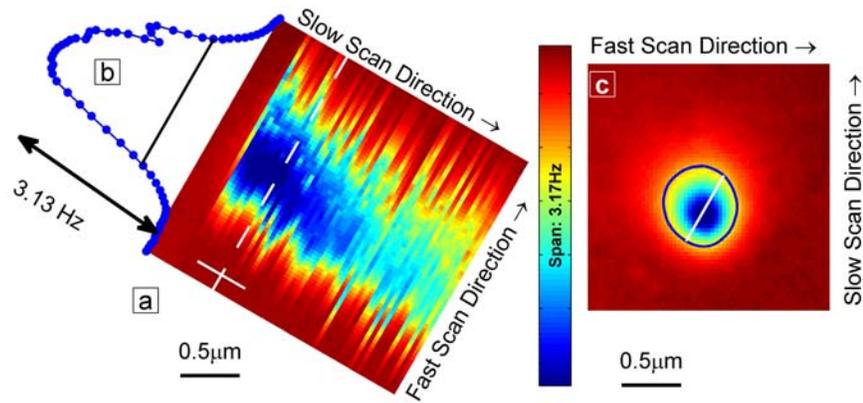

**Supplementary Figure 3:** Extracting $w(\varphi)$ for Fig. 4 in the main text. (A) Scan at $T = 20$K at height z=80nm ( $F_{lat}^{max} \approx 20$pN) at angle 60° to $\hat{x}$. Dashed line denotes scan line cutting region of extremal vortex signal. Solid line denotes range from which $w(\varphi)$ was determined. (B) Single line scan extracted at the dashed line. Line denotes 0.23 of the peak value, where we extract the width. (C) Scan at $T = 5.2$K. The width of the static vortex is extracted by the same procedure as in panels A and B.

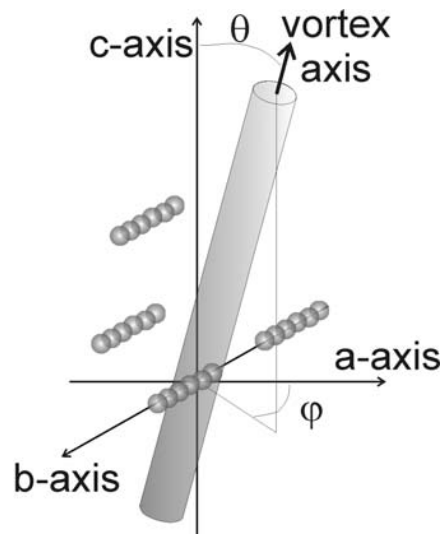

**Supplementary Figure 4:** Cartoon showing crystal axes, vortex axis, the clusters and the angles used in the equations, $\theta$ and $\varphi$.

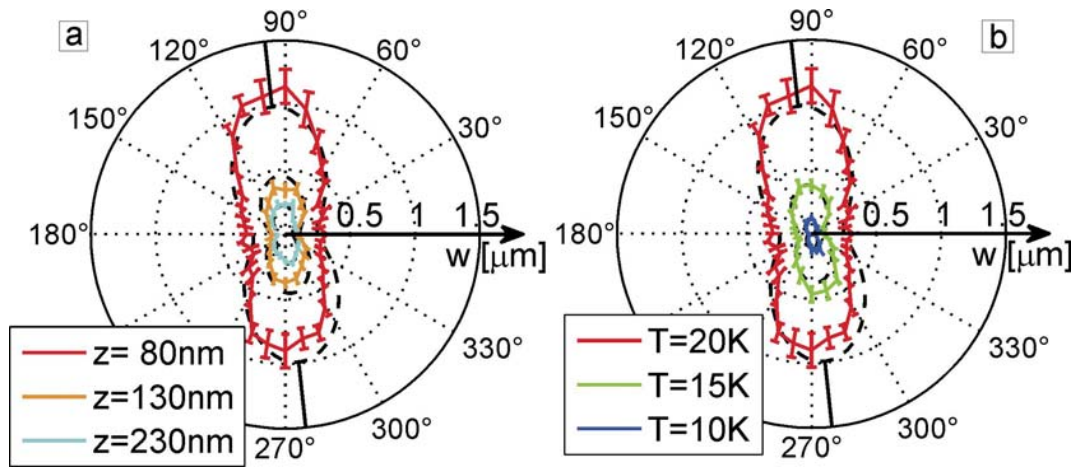

**Supplementary Figure 5:** Distance moved, $w$, measured at the maximum lateral force, versus scan angle $\varphi$. Bars indicate 70% confidence intervals. Dashed lines show fit to the model, assuming no correlations in the pinning, as explained in Supplementary Discussion 4. For the fit we constrain the same $\zeta$ for all data sets. Additional parameters are an overall scale factor for each data set and the angle between $\hat{y}$ and the b-axis, $\varphi_0$. Fit results: $\zeta = 1.6$ and $\varphi_0 = 6°$. The solid lines indicate the b-axis from the fit. **(A)** $\varphi$-dependence at $T = 20$K as a function of scan height, which controls the magnitude of the force (for 80nm, 130nm, 230nm: $F_{lat}^{max} \approx 20$, 15, 10pN). **(B)** $\varphi$-dependence, for scan height 80nm, as a function of temperature.

## Supplementary Table – Cantilevers[(1)]

$d$ = nominal thickness of Fe coating, $f_0$ = resonant frequency, $c$ = cantilever spring constant,[(2)] $Q_{ambient}$ = quality factor for the oscillations measured in ambient conditions, $Q_{vacuum}$ = quality factor for the oscillations measured at 80K at pressure $<10^{-6}$ mbar. $\widetilde{m}$ , $h_0$ = fit parameters for the monopole-monopole model, explained in Supplementary Discussion 1.

| Tip name | Relevant figures | $d$ [nm] | $f_0$ [kHz] | $c$ [(2)] [N/m] | $Q_{ambient}$ | $Q_{vacuum}$ | $\widetilde{m}$ [(3)] [nAm] | $h_0$ [(3)] [nm] |
|---|---|---|---|---|---|---|---|---|
| A | Figs. 2-4, inset to Fig. 1d Supplementary Figs. 2,3,5 | 60 | 51 | 2.3±0.2 | 350 | >4E5 | 32±2 | 360±10 |
| B[(4)] | Figs. 1 (main panels) Supplementary Fig. 1[(3)] | 60 | 75 | 4.0±0.1 | 300 | >4E5 | 29±6 | 360±40 |

[(1)]**Cantilevers used**. We used commercially available silicon cantilevers.[7] Each was electron beam deposition coated by an iron film.

[(2)]**Determining the spring constant.** We used Sader's method.[8]

[(3)]**Fit parameters.** $\widetilde{m}$ and $h_0$ change every time the tip magnetization changes. The numbers quoted here are for the runs from which we generated the listed figures, with the exception of Supplementary Figure 1, for which we did not determine $\widetilde{m}$ and $h_0$ .

[(4)]**In-plane tip orientation.** To rule out the possibility that the apparent a-b anisotropy was related to a residual in-plane component of the tip magnetization, we mounted tip B at an angle with respect to tip A, to preserve the scan directions while ensuring that any in-plane magnetization component was rotated.

**Supplementary Discussion 1: Monopole-monopole model for tip-vortex interaction**

The simplest model for tip-vortex interaction is the monopole-monopole model, so called because the interaction is identical to the interaction between magnetic monopoles. This model is relevant under two assumptions:

1. The tip is an infinitely long and narrow cylinder with axis and magnetization along $\hat{z}$.

2. $z >> \lambda_{ab}$.

When $z >> \lambda_{ab}$, the field from a vortex is well approximated by the field from a monopole residing $\lambda_{ab}$ below the surface of the superconductor, which is assumed to fill the half space $z \leq 0$ :[9, 10]

$$\vec{B}\left[\vec{R}, z\right] \approx \frac{\Phi_0}{2\pi} \frac{\left(\vec{R} + (z + \lambda_{ab})\hat{z}\right)}{\left(R^2 + (z + \lambda_{ab})^2\right)^{3/2}} . \qquad (S1)$$

The resulting force acting on the tip due to the interaction with the vortex is $\vec{F}\left(\vec{R}, z\right) \approx \tilde{m}\vec{B}\left(\vec{R}, z\right)$, where $\tilde{m}$ is the dipole moment per unit length of the tip.

In the main text in we find that the amplitude of the signal from vortices in the YBCO single crystal is well described by one temperature independent curve. Moreover, we find good fit with the monopole-monopole model:

$$\max\left(\partial F_z / \partial z\right) = \left(\tilde{m}\Phi_0 / \pi\right) / \left(z + h_0\right)^3 . \qquad (S2)$$

Here $h_0 = \lambda_{ab} + \Delta z + \delta \tilde{z}$, where $\Delta z$ is any difference between the bottom of the tip and the location of the magnetic coating and $\delta \tilde{z}$ is an estimate for the thickness of the dead layer. The dead layer is a non-superconducting layer at the surface of the superconductor,[11] which is known to develop over time when the sample is in storage. The parameters of the fit in Fig. 3a in the main text are listed in the Supplementary Table. The values of $\tilde{m}$, the tip magnetic moment per unit length, are consistent with our tip parameters. In particular, dividing by the typical magnetization of iron, $M_0^{Fe} \approx 1.7 \cdot 10^6$ A/m,[12] we find a cross-section near $2 \cdot 10^4$ nm$^2$. Dividing by $d$, the Fe layer thickness, this gives a tip radius on the order of 100 nm. Finally, noting that $\lambda_{ab} \approx$ 100 nm, $h_0 \approx 350$ nm gives a dead layer thickness on the scale of 100 nm. This large scale for the dead layer is consistent with the sample being held in storage for a few years at room temperature.

**Supplementary Discussion 2: Weak collective pinning in biaxial superconductors.**

Weak collective pinning theory describes how a vortex interacts with randomly distributed weak pinning centres, which can pin a vortex only by cumulative effect. It allows characterization of the way a vortex meanders through the random pinning potential. Besides externally applied forces, this is the result of the interplay between pinning and elasticity. On one hand, the more a vortex can bend the better advantage it can take of pinning. On the other hand, bending increases the elastic energy. As a result the vortex will only bend on large length scales and will behave rigidly on shorter length scales.

When the dominant source of pinning are randomly distributed weak point pinning sites, the typical pinning energy for an object taking up a volume $V$ is $U_{pin}\sqrt{n_i V}$, where $n_i$ is the pinning site density and $U_{pin}$ is the energetic price of depinning. The square root results from the randomness of the meander of the vortex through the pinning potential. It is akin to a random walk. The effective volume taken up by a vortex segment is roughly $V = \xi_a \xi_b L$, where $L$ is the length and $\xi_a$ and $\xi_b$ are the radii of the core along the a and b axes, so that the energy per unit length is:

$$\sqrt{\left\langle \varepsilon_{pin}^2 \right\rangle} \Big/ L \approx U_{pin} \sqrt{n_i \xi_a \xi_b / L} \; . \tag{S3}$$

This energy is balanced against the elastic energy per unit length, $\frac{1}{2}\varepsilon_\perp\left(u/L\right)^2$, where $u$ is the displacement of the tilted vortex from its non-tilted position, and $\varepsilon_\perp$ is the tilt modulus. We estimate $L_c$ by identifying $u$ with the typical scale of the fluctuations in the disorder.

When the vortex tilts at an azimuthal angle $\varphi$, we set $u = \xi(\zeta, \varphi)$, where $\xi(\zeta, \varphi) = \xi_{ab}\sqrt{\eta(\zeta, \varphi + \pi/2)}$, $\xi_{ab} = \sqrt{\xi_a\xi_b}$ and $\eta(\zeta, \varphi) \equiv \zeta\cos^2\varphi + \zeta^{-1}\sin^2\varphi$. Optimizing $L$ to minimize the free energy per unit length, $\frac{1}{2}\varepsilon_\perp(\zeta, \varphi)\left(\xi(\zeta, \varphi)/L\right)^2 - U_{pin}\sqrt{n_i\xi_{ab}^2/L}$, we obtain:

$$L_c(\zeta, \varphi)/L_c^0 = \left[\eta(\zeta, \varphi)\eta(\zeta, \pi/2 + \varphi)\right]^{2/3} \qquad \text{(S4)}$$

where $L_c^0 = \left(2\varepsilon_0\varepsilon^2\sqrt{\xi_{ab}^2/n_i}\Big/U_{pin}\right)^{2/3}$ is the result for a uniaxial superconductor.[13] Here we assume that even after it tilts, the vortex is nearly parallel to the c-axis. Each segment of length $L_c$ is pinned with force up to $F_p$, which is given by:

$$F_p \approx \sqrt{\left\langle \varepsilon_{pin}^2 \right\rangle}\Big/\xi(\zeta, \varphi) = U_{pin}\sqrt{n_i\xi_{ab}^2L_c}\Big/\xi(\zeta, \varphi). \qquad \text{(S5)}$$

**Supplementary Discussion 3: Total displacement versus applied force.**

In the model, we assume that the vortex breaks into segments of length $L_c$. Each segment interacts with its neighbors on both ends by an elastic force, which we model as a spring with spring constant $k$. We assume that when the vortex is dragged the only distortion is stretching along the drag direction. We denote by $u_n$ the displacement of the $n^{\text{th}}$ segment from the $(n+1)^{\text{th}}$ segment, where n enumerates the segments starting from the surface. For each segment, we have:

$$F_n - ku_n + ku_{n-1} = F_p, \qquad \text{(S6)}$$

with $u_N = 0$ and $u_0 \equiv 0$. Assuming the dragging force on each segment decays exponentially: $F_n = F_1\exp\left(-(n-1)L_c/\Lambda\right)$. Summing Eq. S6 over $n$:

$$F = \sum_{n=1}^{N} F_n = F_1 \sum_{n=1}^{N} q^{n-1} = F_1 \frac{1-q^N}{1-q} = N F_p \, , \qquad (S7)$$

where $q = \exp(-L_c / \Lambda)$ and the total external force on the $N$ segments that move is $F$. Multiplying Eq. S6 by $n$ and summing gives $\sum_{n=1}^{N} n F_n + k \sum_{n=1}^{N} u_n = \sum_{n=1}^{N} F_p$. Noting that $w = \sum_{n=1}^{N} u_n$, we find:

$$w = \frac{1}{k} \left( \frac{N(N+1)}{2} F_p - F_1 \left( \frac{1 - (N+1)q^N}{1-q} + \frac{q(1-q^N)}{(1-q)^2} \right) \right). \qquad (S8)$$

Using the last equality in Eq. S7, this gives:

$$w \approx \frac{F}{2kF_p} \left( -F + \frac{2}{1-q^{F/F_p}} F - \left( \frac{1+q}{1-q} \right) F_p \right), \qquad (S9)$$

The approximation in Eq. S9 is $F/F_p \approx N$, which is appropriate for $F/F_p \gg 1$. The elastic constant $k$ accounts for the resistance of the vortex to bending. The behaviour of Eq. S9 for $F \gg F_p \Lambda / L_c$ gives the behaviour quoted in the main text.

**Calculation of $k$.** We start from the energy per unit length of a vortex in a biaxial crystal, which can be shown to be given by $\varepsilon_l(\varphi, \theta) = \varepsilon_0 \sqrt{\varepsilon^2 \eta(\zeta, \varphi) \sin^2 \theta + \cos^2 \theta}$,[14] where $\varphi$ and $\theta$ are the azimuthal and polar angles the vortex is pointing along, as defined in Supplementary Figure 4, and $\varepsilon \equiv \lambda_{ab} / \lambda_c$. We defined $\eta(\zeta, \varphi)$ above and $\varepsilon_0 = \frac{(\Phi_0 / \lambda_{ab})^2}{4\pi\mu_0} \log(\lambda_{ab} / \xi_{ab})$ is the line tension.

Next we write the total elastic energy of the stretched vortex, not including pinning, as a sum over all segments:

$$E_{elastic} = \sum_{n=1}^{N} \varepsilon_l(\varphi, \theta_n) L_c / \cos\theta_n \, , \qquad (S10)$$

where $\theta_n$ is the polar angle of segment $n$. Assuming $\theta_n \ll 1$, we expand to second order in $\theta_n$ and find that, up to a constant:

$$E_{elastic} \sim \sum_{n=1}^{N} \tfrac{1}{2}\, \varepsilon_\perp\left(\zeta,\varphi\right) L_c \theta_n^2 \; , \tag{S11}$$

where $\varepsilon_\perp\left(\zeta,\varphi\right)=\varepsilon_0\varepsilon^2\eta\left(\zeta,\varphi\right)$. Next we note that $\theta_n \sim u_n/L_c$ for $\theta_n <<1$, and write $E_{elastic}$ in terms of $u_n$. Finally, we take a derivative in $u_n$ and obtain the second two terms on the left hand side of Eq. S6, with the identification:

$$k \equiv \varepsilon_\perp/L_c \; . \tag{S12}$$

**Supplementary Discussion 4: Polar dependence of the dragging distance without correlations in pinning.**

According to Eq. S9 (and Eq. 1 in the main text): $w \propto L_c/\varepsilon_\perp F_p$. This implies a plane-anisotropic $w$, as follows. The penetration depth anisotropy contributes through $\varepsilon_\perp$ because[13, 14] $\varepsilon_\perp\left(\zeta,\varphi\right) \propto \eta\left(\zeta,\varphi\right)$, where $\eta\left(\zeta,\varphi\right)$ was defined above, implying that it is easier to tilt a vortex towards the b-axis. Additional in-plane anisotropy is from the interaction between the vortex core and pinning sites. Even for an isotropic pinning site, the effective pinning potential is steepest for $\varphi=0$ because of the in-plane anisotropy of the vortex core radius, $\xi\left(\zeta,\varphi\right)$. In GL $\xi\left(\zeta,\varphi\right)$ is proportional to the usual coherence length,[13, 15] implying $\xi\left(\zeta,\varphi\right)=\xi_{ab}\sqrt{\eta\left(\zeta,\varphi+\pi/2\right)}$ and a larger depinning force along the a-axis. Finally, using $\varepsilon_\perp\left(\zeta,\varphi\right) \propto \eta\left(\zeta,\varphi\right)$, $L_c\left(\zeta,\varphi\right) \propto \left[\eta\left(\zeta,\varphi\right)\eta\left(\zeta,\varphi+\pi/2\right)\right]^{2/3}$ and $F_p \propto \sqrt{L_c\left(\zeta,\varphi\right)}/\xi\left(\zeta,\varphi\right)$, we find $w\left(\zeta,\varphi\right) \propto \eta^{5/6}\left(\zeta,\varphi+\pi/2\right)/\eta^{2/3}\left(\zeta,\varphi\right)$. We use this expression to fit the data in Figs. 4c,d in the main text. The result, as well as the raw data, is presented in SFig. 5. We find that the angle between the a-axis and $\hat{x}$ is $\varphi_0=6°$, consistent with the conclusion from Supplementary Figure 1, and that $\zeta=1.6$. As mentioned in the main text this value for the a-b anisotropy is too large.

**Supplementary Discussion 5: Effect of correlations in the disorder on the polar dependence of the dragging distance**

The analysis above presupposes that the positions of the point pinning sites are completely uncorrelated. Formally, this enters the theory through a correlation function, which for the purely random case is given by: $\left\langle U_{pin}(\vec{r}) U_{pin}(\vec{r}') \right\rangle \propto \delta(\vec{r} - \vec{r}')$, in which $U_{pin}$ is the pinning potential for a single pinning site. In YBCO the vacancies, which are the point pinning sites in the model, are known to form clusters along the Cu-O chains along the b-axis.[16, 17] We account for that by endowing the $\delta$-function with a finite width along the b-axis. For concreteness and tractability we choose:

$$\left\langle U_{pin}(\vec{r}) U_{pin}(\vec{r}') \right\rangle \propto \delta(x - x') \frac{\pi^{-1} R_b}{(y - y')^2 + R_b^2} \delta(z - z'), \qquad \text{(S13)}$$

in which $R_b$ is a length scale that characterizes the correlation length of the disorder along the Cu-O chains. In Eq. S13 is the implicit assumption $R_b > 0$. Here we chose a coordinate system in which $\hat{x}$ is parallel to the a-axis, $\hat{y}$ to the b-axis and $\hat{z}$ to the c-axis. Assuming $\delta T_c$-disorder,[13] assuming $R_b >> \xi_{ab}$ and using Eq. S13 we find that the energy per unit length is given by Eq. S3 multiplied by an additional angle dependent factor:

$$\beta(\zeta, \varphi) \equiv \sqrt{\frac{\xi_b \arctan(\sin(\varphi)\xi(\zeta, \varphi)/R_b)}{\sin(\varphi)\xi(\zeta, \varphi)}} \ . \qquad \text{(S14)}$$

This implies, using analysis similar to that leading to Eq. S4, that $L_c$ attains an additional dependence on $\varphi$, with $\gamma(\zeta, \varphi) \equiv \eta^{1/6}(\zeta, \pi/2 + \varphi)/\beta^{2/3}(\zeta, \varphi)$, multiplying the right hand side of Eq. S4. As a result there is an additional factor of $\gamma(\zeta, \varphi)$ dividing both $F_p$ and $k$. The fit using these formulae, shown in Figs. 4c,d in the main text gives the cluster size quoted there and an angle of 6° between the a-axis and $\hat{x}$. We note that the fit gives a value of $R_b$ that is on the same scale as the vortex core size $\xi_{ab}$,[18] and not much larger, as we assumed. We thus take the results of the fit to give only the correct order of magnitude of $R_b$ .